\documentclass[prb,longbibliography,reprint,superscriptaddress,citeautoscript]{revtex4-1}

\usepackage[applemac]{inputenc}
\usepackage[T1]{fontenc}
\usepackage[american]{babel}
\usepackage[scaled]{helvet}
\usepackage{graphicx,amssymb,bm,mathtools,textcomp,courier,multirow,multirow,bigdelim,color}
\usepackage[normalem]{ulem}	

\newcommand{\Dirac}[3]{\left\langle #1 \left| #2\right| #3\right\rangle}

\newcommand{\product}[2]{\left\langle #1 \left| #2 \right\rangle\right.}

\newcommand{\ket}[1]{\left|\left. #1 \right\rangle\right.}

\newcommand{\cre}[1]{\hat{c}^{\dagger}_{#1}}
\newcommand{\ann}[1]{\hat{c}_{#1}}
\newcommand{\abs}[1]{\left| #1 \right|}
\newcommand{\av}[1]{\left\langle #1 \right\rangle}

\newcommand{\figureshortname}{Fig.}
\newcommand{\equationshortname}{Eq.}

\newcommand{\eref}[1]{\equationshortname~\eqref{#1}}
\newcommand{\sref}[1]{Sec.~\ref{#1}}
\newcommand{\aref}[1]{Appx.~\ref{#1}}
\newcommand{\cref}[1]{Chapter~\ref{#1}}
\newcommand{\fref}[1]{\figureshortname~\ref{#1}}
\newcommand{\tref}[1]{\tablename~\ref{#1}}
\newcommand{\rcite}[1]{Ref.~[\onlinecite{#1}]}

\newcommand{\mmrcite}[3]{Refs.~[\onlinecite{#1},\onlinecite{#2},\onlinecite{#3}]}

\graphicspath{{Pictures/}}

\begin{document}
\def\sectionautorefname{Sec.}

\title{Validity of the single-particle description and charge noise resilience for multielectron quantum dots}

\author{Michiel A. Bakker}
\thanks{These two authors contributed equally to this work}
\affiliation{Peter Grünberg Institute (PGI-2), Forschungszentrum Jülich, D-52425 Jülich, Germany}
\affiliation{Delft University of Technology, PO Box 5046, 2600 GA Delft, The Netherlands}

\author{Sebastian Mehl}
\thanks{These two authors contributed equally to this work}
\email{s.mehl@fz-juelich.de}
\affiliation{JARA-Institute for Quantum Information, RWTH Aachen University, D-52056 Aachen, Germany}
\affiliation{Peter Grünberg Institute (PGI-2), Forschungszentrum Jülich, D-52425 Jülich, Germany}

\author{Tuukka Hiltunen}
\author{Ari Harju}
\affiliation{COMP Centre of Excellence, Department of Applied Physics, Aalto University, Helsinki, Finland}

\author{David P. DiVincenzo}
\affiliation{JARA-Institute for Quantum Information, RWTH Aachen University, D-52056 Aachen, Germany}
\affiliation{Peter Grünberg Institute (PGI-2), Forschungszentrum Jülich, D-52425 Jülich, Germany}

\date{\today}

\begin{abstract}
We construct an optimal set of single-particle states for few-electron quantum dots (QDs) using the method of natural orbitals (NOs). The NOs include also the effects of the Coulomb repulsion between electrons. We find that they agree well with the noniteracting orbitals for GaAs QDs of realistic parameters, while the Coulomb interactions only rescale the radius of the NOs compared to the noninteracting case. We use NOs to show that four-electron QDs are less susceptible to charge noise than their two-electron counterparts.
\end{abstract}

\maketitle

\section{Introduction}

Quantum dots (QDs) are zero dimensional quantum systems with many characteristic properties of atoms;\cite{jacak1998,kouwenhoven2001,reimann2002} for example, an energy shell structure was found using transport measurements \cite{reed1988,reed1993,ferry2009}. We are interested in the potential to store quantum information \cite{nielsen2000} using the electron spin of QDs \cite{loss1998,awschalom2002}. We consider the electron configurations of a gate-defined QD with the lowest energies as a realization of a quantum memory. Specifically we discuss a two-electron configuration of a QD, where the low-energy properties are described by a singlet state and the triplet states. The $s_z=0$ subspace is two dimensional and defines a qubit \cite{divincenzo2000-2}, which we call singlet-triplet qubit (STQ) in the following \cite{levy2002}.

STQs realize universal quantum computation for an array of QDs when the transfer of electrons between neighboring QDs is permitted \cite{levy2002,taylor2005,taylor2007}. Many experiments have shown that high-fidelity quantum gates can be constructed for STQs encoded using gate-defined QDs.\cite{foletti2009,bluhm2010,maune2012,wu2014,veldhorst2014,veldhorst2014-2} In the following, we only consider GaAs QDs, which have weak spin-orbit interactions such that the spin and the orbital parts of QD wave functions are decoupled. Since we only work with the $s_z=0$ subspace, we are able to neglect the Zeeman interaction and always consider the $s_z=0$ triplet without further distinguishing it from other triplet states.

The Coulomb interactions between electrons, and consequently correlation effects for QD electrons, are more important compared to electrons bound to atoms because the sizes of the QD wave functions are larger than atomic wave functions \cite{ashoori1996}. Early descriptions of QDs assumed that Coulomb interactions provide a large energy contribution to each electron that is added to a QD, but the magnitude of this energy contribution is independent of the electron configuration \cite{averin1986,averin1991}. Very often such a description is insufficient because correlations can induce novel effects in the energy spectrum of QDs \cite{hawrylak1993,guclu2008}. Even more, it has been pointed out that correlations can provide all the necessary gate operations for universal quantum computation \cite{kyriakidis2002,kyriakidis2005-1,kyriakidis2007}.

Weakly correlated quantum systems can be described in a mean-field approximation, as for Hartree-Fock calculations (cf., e.g., \rcite{reimann2002}). Mean-field calculations are simple, and also the wave functions of multielectron QDs can be constructed \cite{reusch2001,reusch2003}. Hartree-Fock wave functions are, however, restricted to a single Slater determinant, and a Hartree-Fock description usually fails to correctly describe strongly correlated quantum systems. Large correlations are expected for weakly confined QDs \cite{johnson1992}. Even for realistic QD parameters, numerical calculations proved the failure of Hartree-Fock calculations.\cite{pfannkuche1993}

Computationally powerful methods have been developed to describe the correlation effects of few-electron QDs \cite{rontani2006}. Very often a full-configuration interaction (full-CI) method is used to analyze QD wave functions \cite{helgaker2000}. Full-CI calculations diagonalize the QD Hamiltonian using a large subspace of possible QD eigenfunctions. The predicted eigenfunctions of the QDs are exact in the chosen basis, but a highly accurate prediction of the ground state wave function requires a large set of basis states. We show that interaction effects for QDs can be described more efficiently using an optimized basis set for CI calculations. Löwdin described a method to construct effective orbital wave functions for an interaction quantum system, which are called the natural orbitals (NOs) \cite{lowdin1955-1,*lowdin1955-2,*lowdin1955-3}. It is well known in quantum chemistry that these orbitals are an optimal basis for CI calculations \cite{szabo1996,mcweeny1992}.

We show that few-electron QDs can be described very efficiently using Löwdin's NOs. Starting from a full-CI analysis of QDs, we derive the NOs for realistic GaAs QDs. We find that the NOs with the highest occupations have a very transparent interpretation because they resemble the noninteracting eigenstates of QDs that have been rescaled as a consequence of Coulomb interactions. The explicit constructions of NOs will also provide an analytical description of few-electron QDs. The NO description is a comprehensive tool to analyze QDs without disregarding correlation effects, and we point out that NOs can be constructed independently from a full-CI calculation.\cite{mcweeny1992,szabo1996}

\begin{figure}
\centering
\includegraphics[width=0.49\textwidth]{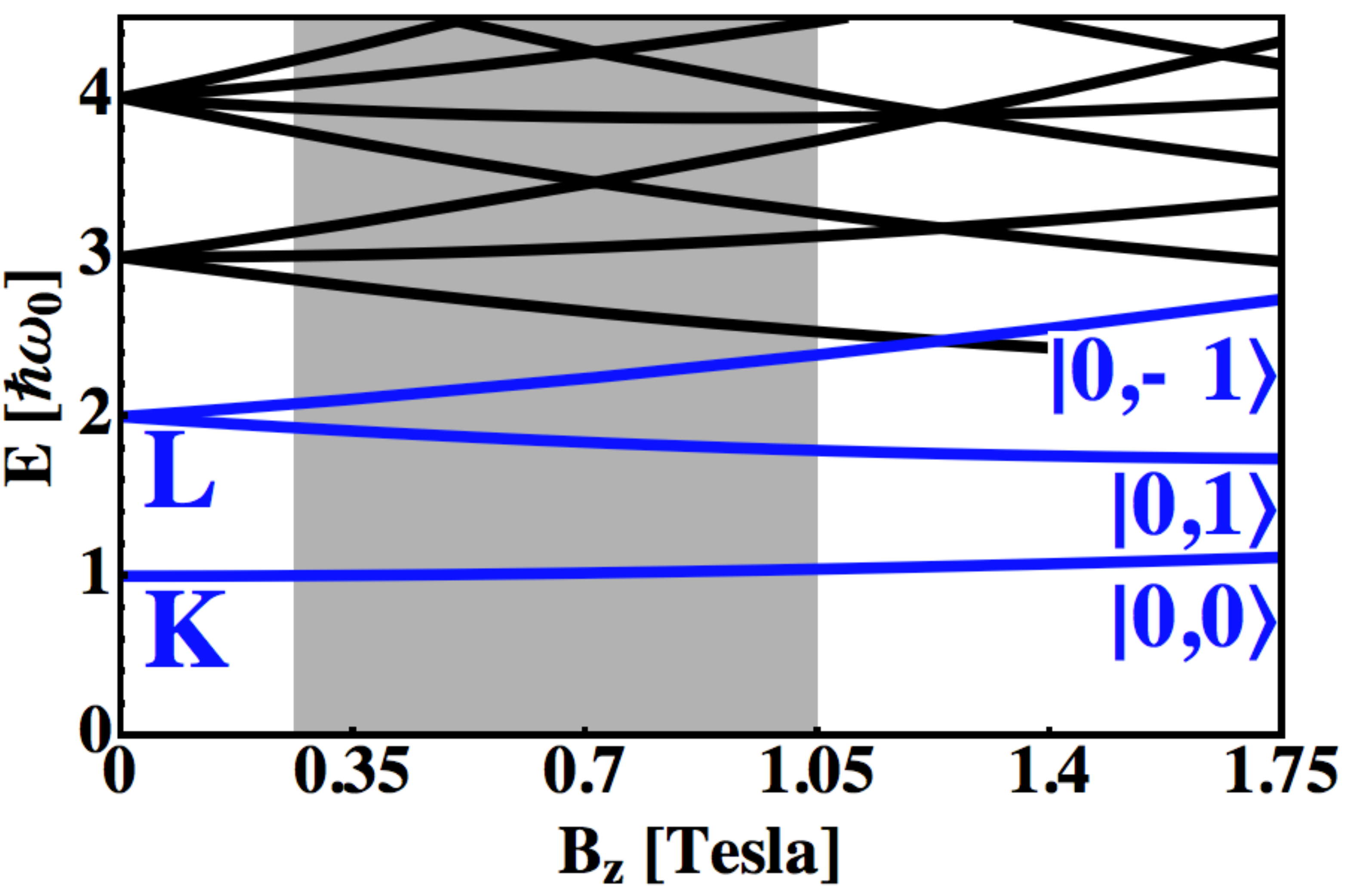}
\caption{
Noninteracting orbital energy spectrum of a harmonic QD for the confining strength $\hbar\omega_0=3~\text{meV}$ and out-of-plane magnetic fields $B_z$ according to \eref{eq:HAMILTONIAN}. The eigenstates are the FD states $\ket{n,l}$ that are described by the orbital quantum number $n$ and the magnetic quantum number $l$. For $B_z=0$, the FD states are grouped into atomic shells. We label the nondegenerate lowest shell as K shell and the doubly degenerate first-excited shell as L shell. Out-of-plane magnetic fields lift the degeneracies. The QD parameters from the highlighted region protect a four-electron STQ from charge noise, as described in the main text.
\label{fig:01}}
\end{figure}
 
Finally, the NO description is applied to study the coherence properties of STQs. We focus on charge noise because fluctuating electric potentials directly couple to the qubit's wave functions and are one of the limiting obstacles to realizing a quantum memory \cite{hanson2007-2,taylor2007,hu2006}. \rcite{mehl2013} predicted a perfect insensitivity to charge noise of a STQ that is encoded using a four-electron QD.\footnote{
\rcite{mehl2013} suggests a qubit encoding in a six-electron double QD because the (4,2) configuration is well protected from charge noise. In this case, the four-electron QD determines the sensitivity to charge noise. Note the related proposal of a six-electron double QD in \rcite{nielsen2013}.
} The noise protection was derived from a shell filling model for noninteracting QD electrons \cite{kouwenhoven2001}. We find that the four-electron configuration remains noise insensitive even in the presence of Coulomb interactions because the NOs match very closely to noninteracting QD eigenstates. When Coulomb interactions are included in the description of QD electrons, we find nearly an order of magnitude increase in coherence for a four-electron STQ compared to a two-electron STQ.

The organization of the paper is as follows. \sref{sec:NO} introduces a model to describe few-electron QDs using the noninteracting eigenfunctions of a QD. This model predicts a high protection from charge noise for STQs encoded using four-electron QDs. \sref{sec:Int} analyzes interacting QD electrons using NOs, which confirms the validity of the noninteracting shell filling model. We also characterize the use of NOs in CI calculations. \sref{sec:Noise} describes the coherence properties of few electron QDs under charge noise. We find that a four-electron STQ is better protected from charge noise than a two-electron STQ, as predicted from noninteracting QD electrons. \sref{sec:Sum} summarizes the findings of the paper.

\section{
\label{sec:NO}
Description of Noninteracting Few-Electron Quantum Dots}

A starting description of few-electron QDs is obtained using the noninteracting eigenstates of a QD. We consider the single-particle Hamiltonian
\begin{align}
h\left(\bm{r}\right)=
-\frac{\hbar^2}{2m}\bm{\nabla}^2+
\frac{m\left[\omega_0^2+\left(\frac{\omega_c}{2}\right)^2\right]}{2}\left(x^2+y^2\right)-
\frac{\omega_c}{2} l_z
\label{eq:HAMILTONIAN}
\end{align}
of a harmonic confining potential in two dimensions with the strength $\hbar\omega_0$ and $\bm{r}=\left(x,y\right)^T$. The orbital contributions of a magnetic field are introduced through the vector potential in the symmetric gauge $\bm{A}=\frac{B_z}{2}\left(y,-x,0\right)^T$. The out-of-plane magnetic field appears in the QD Hamiltonian of \eref{eq:HAMILTONIAN} through the cyclotron frequency $\omega_c=\frac{eB_z}{m}$. It gives an additional contribution to the harmonic confinement in the second term of \eref{eq:HAMILTONIAN}. Additionally it distinguishes the motion of clockwise and counterclockwise circulating electrons through the orbital moment $l_z=\frac{\hbar}{i}\left(x\partial_y-y\partial_x\right)$.

The eigenvectors of \eref{eq:HAMILTONIAN} are the Fock-Darwin states (FD states) \cite{fock1928,darwin1931}
\begin{align}
\label{eq:FD}
\psi_{nl}\left(\bm{r}\right)=&
\sqrt{\frac{n!}{\pi\left(n+l\right)!}}
\left(\frac{m\Omega}{\hbar}\right)^{\frac{\abs{l}+1}{2}}
L_n^l\left(\frac{m\Omega}{\hbar}\abs{\bm{r}}^2\right)
\\\nonumber &
\times e^{-il\arg\left(\bm{r}\right)}\abs{\bm{r}}^{\abs{l}}
e^{-\frac{m\Omega}{2\hbar}\abs{\bm{r}}^2},
\end{align}
with $\Omega=\sqrt{\omega_0^2+\left(\frac{\omega_c}{2}\right)^2}$. $L_n^l\left(x\right)$ are the generalized Laguerre polynomials. The FD states are described by the orbital quantum number $n$ and the magnetic quantum number $l$.\cite{jacak1998}

As in atomic physics, $\psi_{nl}\left(\bm{r}\right)$ are grouped into orbital shells. The eigenenergies of $\psi_{nl}\left(\bm{r}\right)$ are $E_{nl}=\left(2n+\abs{l}+1\right)\hbar\Omega-l\frac{\hbar\omega_c}{2}$. \fref{fig:01} shows the eigenenergies of the FD states as a function of the out-of-plane magnetic field $B_z$ for a GaAs QD with a typical confining strength, $\hbar\omega_0=3~\text{meV}$ [\onlinecite{burkard1999}], and a QD electron with the effective mass $m=0.067m_e$, where $m_e$ is the electron mass. In the absence of magnetic fields, a highly regular level spectrum is observed, where we label the orbitally nondegenerate ground state as K shell and the doubly-degenerate first-excited states as L shell. The orbital components of the out-of-plane magnetic fields lift all the degeneracies in the energy spectrum, and there is a crossing of energy levels from different energy shells at even higher magnetic fields. In the case of very high magnetic fields (several tesla), the quantum Hall regime is approached with the well-known interpretation of the FD states as Landau levels.

\begin{figure*}
\centering
\includegraphics[width=0.49\textwidth]{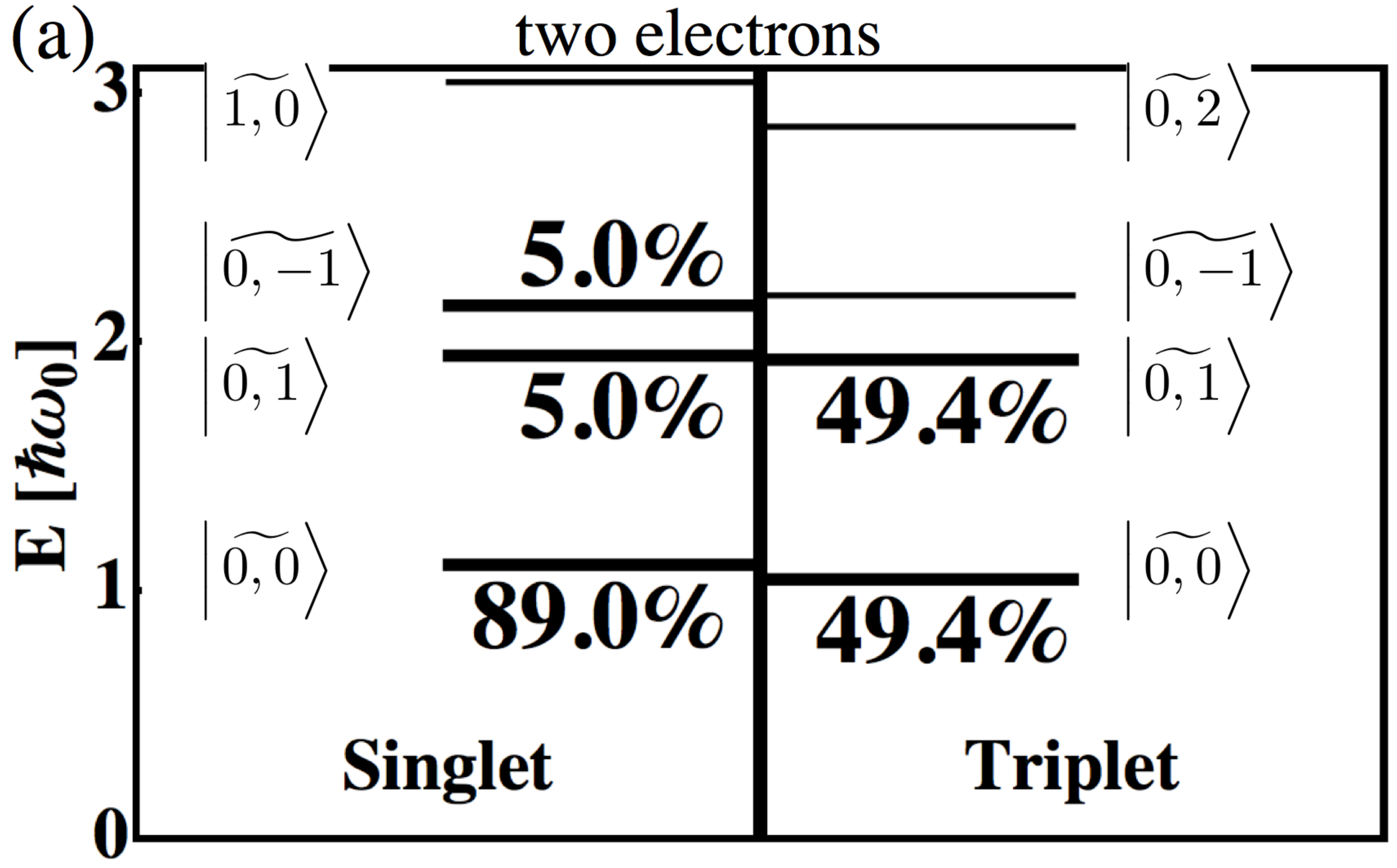}\hfill
\includegraphics[width=0.49\textwidth]{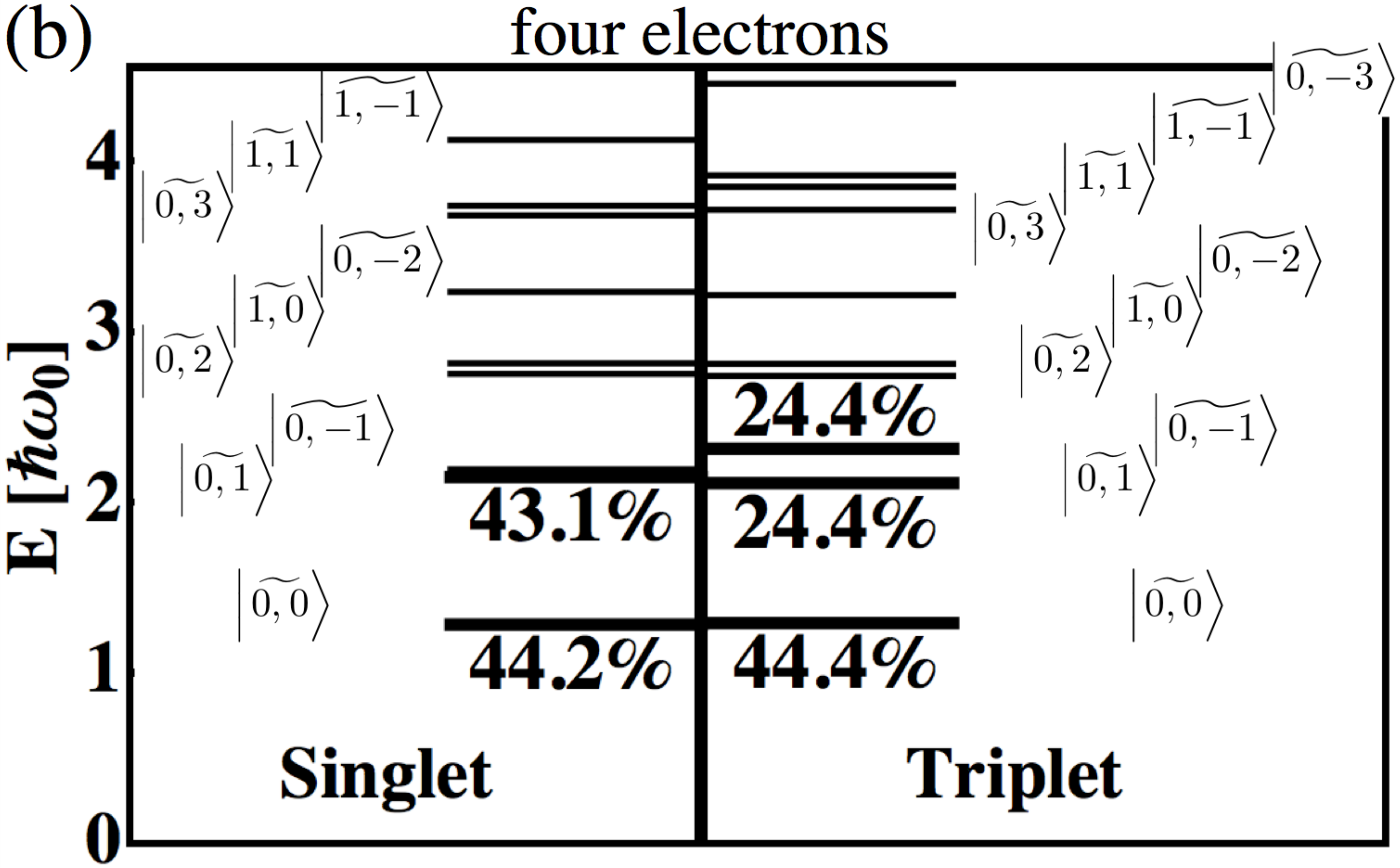}
\caption{
\label{fig:02}
NOs of the lowest energy configurations with $s_z=0$ of a two-electron QD and a four-electron QD of $\hbar\omega_0=3~\text{meV}$ and $B_z=0.35~\text{T}$. In each case, the ground state is a singlet, and the excited state is a triplet. The NOs are ordered by the energy expectation values of the single-particle Hamiltonian from \eref{eq:HAMILTONIAN}: $E=\Dirac{\psi\left(\bm{r}_1,\dots\bm{r}_N\right)}{h\left(\bm{r}_1\right)}{\psi\left(\bm{r}_1,\dots\bm{r}_N\right)}$. We interpret all the NOs by rescaled FD states $\ket{\widetilde{n,l}}$. The dominant NOs (drawn with bold lines) resemble the predictions from a shell filling model: (a) the two-electron singlet has an occupation of $89\%$ in the lowest NO. Two additional NOs with $5.0\%$ occupations are needed to sufficiently describe the singlet. The two-electron triplet has nearly $99\%$ occupation in the lowest two NOs. All NOs are summarized in \tref{tab:01}. (b) The four-electron configurations have nearly $50\%$ occupations in the ground state orbital. We interpret this configuration as a frozen core. The singlet has the dominant remaining occupation ($43.1\%$) in the next excited NO; the triplet requires two additional NOs (each with $24.4\%$ occupation). In every case, some additional NOs have occupations of a few percent or less; all NOs are tabulated in \tref{tab:02}.
}
\end{figure*}

We study the limit of moderate magnetic fields, where all the state degeneracies are lifted, but the L shell has not been crossed by a state from a different shell yet (cf. the highlighted region in \fref{fig:01}). We quickly review the noise protection criterion from \rcite{mehl2013}. In the noninteracting shell filling model of \rcite{mehl2013}, for a two-electron QD the electrons are paired in the same orbital ground state $\ket{n,l}=\ket{0,0}$ in the singlet configuration, while the Pauli exclusion principle forbids this configuration for the triplet. Here, only one electron occupies the orbital ground state $\ket{0,0}$, while the other electron is in the first excited state $\ket{0,1}$. In the case of four electrons, the low-energy properties are again well described with a singlet-triplet configuration. In every case, two of the four electrons are paired in a ``frozen core'' in the K shell, and the remaining two electrons fill the two orbitals from the L shell. In the singlet configuration, both electrons are in $\ket{0,1}$, but in the triplet configuration one electron is placed into $\ket{0,1}$, and the other one in $\ket{0,-1}$.

Charge noise directly couples to the charge densities of the QD orbitals. The coherence of a STQ is lost if the ground state orbital has a different charge density from the excited orbital \cite{mehl2013}. In the two-electron configuration, the charge density of the singlet state differs from the triplet's charge density because $\ket{0,0}$ and $\ket{0,1}$ have different charge densities. For the four-electron STQ, only the charge densities of $\ket{0,1}$ and $\ket{0,-1}$ distinguish the singlet and triplet charge densities because in every case two electrons are paired in a frozen core in $\ket{0,0}$. But since $\ket{0,1}$ and $\ket{0,-1}$ are complex conjugate of each other, both states have the same charge densities. A four-electron STQ will therefore be protected from all the noise sources that couple weakly to the charge configuration of a QD. A more detailed noise analysis, which also includes the effects of Coulomb interactions, will be given in \sref{sec:Noise}.

\section{
\label{sec:Int}
Description of Interacting Few-Electron Quantum Dots}

\begin{table}
\caption{
\label{tab:01}
Dominant NOs of a two-electron QD in the (a) singlet and (b) triplet configurations. We interpret the NOs $\ket{\text{NO}}$ as FD states $\ket{\widetilde{n,l}}$ according to \eref{eq:FD}, where the tilde represents a rescaling of the Bohr radius $a_B=\sqrt{\frac{\hbar}{m\omega_0}}$ by $\widetilde{a_B}=\beta a_B$. The overlap of $\ket{\text{NO}}$ can be maximized from the FD-overlap$=\abs{\product{n,l}{\text{NO}}}^2$ to overlap$_1=\abs{\product{\widetilde{n,l}}{\text{NO}}}^2$ with the given scaling factor $\beta$. We use $\beta=1.23$ to calculate overlap$_2$ for all the NOs, which allows a comparison of the NOs with the highest occupations. (a) The two-electron singlet has dominant occupation in $\ket{\widetilde{0,0}}$, but the states $\ket{\widetilde{0,1}}$ and $\ket{\widetilde{0,-1}}$ are also needed to accurately describe the singlet configuration. (b) The two-electron triplet has by far the dominant occupation in $\ket{\widetilde{0,0}}$ and $\ket{\widetilde{0,1}}$, which resembles two orbitals filled with one electron each.
}
\begin{tabular}{lccccc}
\hline\hline
\multicolumn{6}{c}{(a) two-electron singlet}\\
state & occupation & FD-overlap & $\beta$ & overlap$_1$ & overlap$_2$\\
$\ket{\widetilde{1,0}}$ & $0.8\%$ & $90.1\%$ & $0.96$ & $90.9\%$ & $68.7\%$\\
$\ket{\widetilde{0,-1}}$ & $5.0\%$ & $98.4\%$ & $0.92$ & $99.9\%$ & $85.0\%$\\
$\ket{\widetilde{0,1}}$ & $5.0\%$ & $98.4\%$ & $0.92$ & $99.9\%$ & $85.0\%$\\
$\ket{\widetilde{0,0}}$ & $89.0\%$ & $95.1\%$ & $1.23$ & $99.6\%$ & $99.6\%$\\
\\
\multicolumn{6}{c}{(b) two-electron triplet}\\
state & occupation & FD-overlap & $\beta$ & overlap$_1$ & overlap$_2$\\
$\ket{\widetilde{0,2}}$ & $0.5\%$ & $97.7\%$ & $0.91$ & $99.9\%$ & $78.1\%$\\
$\ket{\widetilde{0,-1}}$ & $0.5\%$ & $96.6\%$ & $0.88$ & $99.9\%$ & $80.8\%$\\
$\ket{\widetilde{0,1}}$ & $49.4\%$ & $99.1\%$ & $1.07$  & $100.0\%$& $96.1\%$\\
$\ket{\widetilde{0,0}}$ & $49.4\%$ & $98.0\%$ & $1.14$ & $99.9\%$ & $99.3\%$\\
\hline
\hline
\end{tabular}
\end{table}

\begin{table}
\caption{
\label{tab:02}
The dominant NOs of a four-electron QD in the (a) singlet and (b) triplet configurations with the same definitions as in \tref{tab:01}. $\beta=1.42$ eases the comparison of the NOs with the highest occupations. In each case, nearly half of the occupation is in $\ket{\widetilde{0,0}}$, which is interpreted as a core of two electrons. (a) In the singlet configuration, by far the dominant remaining occupation is in $\ket{\widetilde{0,1}}$, which agrees with the picture of two paired electrons in that orbital. (b) The triplet configuration has one electrons in $\ket{\widetilde{0,1}}$ ($\approx 25\%$), the other one in $\ket{\widetilde{0,-1}}$ ($\approx 25\%$).
}
\begin{tabular}{lccccc}
\hline\hline
\multicolumn{6}{c}{(a) four-electron singlet}\\
state& occupation & FD-overlap & $\beta$ & overlap$_1$ & overlap$_2$\\
$\ket{\widetilde{1,-1}}$ & $0.1\%$ & $91.7\%$ & $0.97$ & $92.2\%$ & $29.6\%$\\
$\ket{\widetilde{1,1}}$ & $0.5\%$ & $88.2\%$ & $1.07$ & $91.8\%$ & $49.7\%$\\
$\ket{\widetilde{0,3}}$ & $1.6\%$ & $99.1\%$ & $0.95$ & $99.9\%$ & $55.3\%$\\
$\ket{\widetilde{0,-2}}$ & $0.8\%$ & $99.1\%$ & $0.95$ & $100.0\%$ & $62.0\%$\\
$\ket{\widetilde{1,0}}$ & $1.0\%$ & $86.1\%$ & $1.10$ & $90.6\%$ & $66.6\%$\\
$\ket{\widetilde{0,2}}$ & $3.8\%$ & $99.8\%$ & $1.02$ & $100.0\%$ & $72.9\%$\\
$\ket{\widetilde{0,-1}}$ & $4.5\%$ & $95.9\%$ & $1.14$ & $99.6\%$ & $90.0\%$\\
$\ket{\widetilde{0,1}}$ & $43.1\%$ & $88.2\%$ & $1.26$ & $99.5\%$ & $96.6\%$\\
$\ket{\widetilde{0,0}}$ & $44.2\%$ & $86.4\%$ & $1.42$ & $99.4\%$ & $99.4\%$\\
\\
\multicolumn{6}{c}{(b) four-electron triplet}\\
state& occupation & FD-overlap & $\beta$ & overlap$_1$ & overlap$_2$\\
$\ket{\widetilde{0,-3}}$ & $0.3\%$ & $94.0\%$ & $0.88$ & $99.9\%$ & $43.9\%$\\
$\ket{\widetilde{1,-1}}$ & $0.3\%$ & $89.9\%$ & $1.06$ & $91.9\%$ & $46.2\%$\\
$\ket{\widetilde{1,1}}$ & $0.3\%$ & $89.9\%$ & $1.06$ & $91.9\%$ & $46.3\%$\\
$\ket{\widetilde{0,3}}$ & $0.3\%$ & $94.1\%$ & $0.88$ & $99.9\%$ & $43.5\%$\\
$\ket{\widetilde{0,-2}}$ & $2.1\%$ & $100\%$ & $1.00$ & $100\%$ & $70.0\%$\\
$\ket{\widetilde{1,0}}$ & $1.0\%$ & $86.3\%$ & $1.09$ & $89.4\%$ & $63.9\%$\\
$\ket{\widetilde{0,2}}$ & $2.1\%$ & $100\%$ & $1.00$ & $100\%$ & $70.0\%$\\
$\ket{\widetilde{0,-1}}$ & $24.4\%$ & $89.8\%$ & $1.24$ & $99.5\%$ & $95.6\%$\\
$\ket{\widetilde{0,1}}$ & $24.4\%$ & $89.8\%$ & $1.24$ & $99.5\%$ & $95.6\%$\\
$\ket{\widetilde{0,0}}$ & $44.4\%$ & $86.0\%$ & $1.43$ & $99.4\%$ & $99.4\%$\\
\hline
\hline
\end{tabular}
\end{table}

Correlations are important for few-electron QDs, which would call into question any predictions of an orbital model for few-electron QDs that is obtained from the noninteracting QD wave functions. To discuss the validity of the noninteracting model from \sref{sec:NO}, we analyze here the eigenfunctions of the $N$-particle QDs that we obtain from an exact diagonalization of the QD Hamiltonian with Coulomb interactions. We use full-CI calculations to derive the exact eigenfunctions of interacting few-electron QDs. The interacting QD problem is solved in the basis of $50$ noninteracting eigenstates from \eref{eq:FD}. \aref{app:Numerics} gives a detailed description of the numerical procedure. The orbital part of the $N$-particle wave function $\psi\left(\bm{r}_1,\bm{r}_2,\dotsc, \bm{r}_N\right)$  is used to construct Löwdin's NOs \cite{lowdin1955-1,*lowdin1955-2,*lowdin1955-3}. For this, we calculate the first-order density matrix
\begin{align}
\varrho\left(\bm{r}^\prime,\bm{r}\right)
=&N\int
d\bm{r}_2\cdot\dotsc\cdot d\bm{r}_N
\\\nonumber
&\times
\psi^*\left(\bm{r}^\prime,\bm{r}_2,\dotsc, \bm{r}_N\right)
\psi\left(\bm{r},\bm{r}_2,\dotsc, \bm{r}_N\right),
\end{align}
which describes the charge densities and all the single-particle properties (cf. \aref{app:RedDens} for a description of the method of reduced density matrices). Note that the probability density
\begin{align}
\rho\left(\bm{r}\right)=
\int d\bm{r}_2\cdot \ldots \cdot\bm{r}_N~
\abs{
\psi\left(
\bm{r},\bm{r}_2,\dots,\bm{r}_N
\right)}^2
\label{eq:ProbDens}
\end{align}
is (up to the normalization factor) directly related to the first-order density matrix: $\rho\left(\bm{r}\right)=\frac{1}{N}\varrho\left(\bm{r},\bm{r}\right)$. The probability densities always integrate to 1.

The NOs are the set of normalized, orthogonal functions $\left\{\phi_\alpha\left(\bm{r}\right)\right\}$ that diagonalize the spectral decomposition of the first-order density matrix:
\begin{align}
\varrho\left(\bm{r}^\prime,\bm{r}\right)=
\sum_{\alpha}
\varrho_{\alpha}
\phi_{\alpha}^*\left(\bm{r}^\prime\right)
\phi_{\alpha}\left(\bm{r}\right).
\end{align}
$\varrho_\alpha$ are called the occupations numbers of the NOs. The interpretation as occupation numbers is supported by the sum rule $\sum_\alpha \varrho_\alpha=N$.\cite{lowdin1955-1,*lowdin1955-2,*lowdin1955-3}

We continue with an analysis of the two-electron singlet and triplet configurations of a QD using NOs, as in Löwdin's study of the helium atom \cite{lowdin1956}. In all the following discussions, the NOs are extracted from a full-CI calculation, according to \aref{app:Numerics}, with the parameters $\hbar\omega_0=3~\text{meV}$ and $B_z=0.35~\text{T}$. These parameters describe typical QDs that are used in experiments to realize quantum computation with spin qubits.\cite{burkard1999,kouwenhoven2001,hanson2007-2}
\fref{fig:02}~(a) shows the NOs with the highest occupation numbers for a two-electron QD. We see that the intuitive shell filling model is by far more valid in the triplet configuration. In this case, nearly $99\%$ of the electronic state occupies the two lowest NOs. In the singlet configuration, simple assignment of two electrons to the lowest NO is less valid. Only $89\%$ of the electrons occupy the lowest NO; two additional NOs need to be added to describe the state sufficiently. There is a simple explanation for why the description with a single NO is less valid in the case of the two-electron singlet compared to the triplet. Because the spatial wave function of the singlet state is symmetric, the electrons are always closer together in a doubly occupied singlet configuration than for any triplet configuration. The Coulomb repulsion is higher for a doubly occupied orbital, and a singlet that is restricted to a single NO has a high energy. The singlet energy can be lowered through virtual excitations, which add small occupations to excited NOs.

\begin{table*}
\caption{
\label{tab:03}
Comparison of the energies of the singlet configuration and the triplet configuration for the two-electron QD and the four-electron QD at $\hbar\omega_0=3~\text{meV}$ and $B_z=0.35~\text{T}$ (cf. \aref{app:Numerics} for a description of the numerical calculations). (a) The energy of the two-electron singlet, which is the ground state configuration; (b) the energy of the two-electron triplet state, which is the first excited state of a doubly occupied QD. The CI calculation with the two dominant NOs from \tref{tab:01} is insufficient to accurately reproduce the results from the full-CI calculation. Using the four dominant NOs from \tref{tab:01} gives already a very satisfactory description of the energy configuration. We obtain a similar finding for the four-electron configuration in the (c) singlet and the (d) triplet configurations. The naive description with two (three) NOs from \tref{tab:02} in the singlet (triplet) configuration is far away from the full-CI results; but adding more NOs from \tref{tab:02} increases the accuracy of the restricted CI calculations.
}
\begin{tabular}{cc}
\hline
\hline
\multicolumn{2}{c}{(a) two-electron singlet}\\
Method&Energy [meV]\\
full-CI & $11.1$\\
2 NOs & $11.8$\\
4 NOs & $11.2$\\
5 NOs & $11.2$\\
\hline
\hline
\end{tabular}
\begin{tabular}{cc}
\hline
\hline
\multicolumn{2}{c}{(b) two-electron triplet}\\
Method&Energy [meV]\\
full-CI & $12.1$\\
2 NOs & $12.6$\\
4 NOs & $12.1$\\
6 NOs & $12.1$\\
\hline
\hline
\end{tabular}
\begin{tabular}{cc}
\hline
\hline
\multicolumn{2}{c}{(c) four-electron singlet}\\
Method&Energy [meV]\\
full-CI & $40.3$\\
2 NOs & $43.3$\\
6 NOs & $41.3$\\
9 NOs & $40.6$\\
\hline
\hline
\end{tabular}
\begin{tabular}{cc}
\hline
\hline
\multicolumn{2}{c}{(d) four-electron triplet}\\
Method&Energy [meV]\\
full-CI & $40.5$\\
3 NOs & $43.3$\\
5 NOs & $41.9$\\
9 NOs & $40.8$\\
\hline
\hline
\end{tabular}
\end{table*}

We find that all NOs still closely correspond to FD states $\ket{\widetilde{n,l}}$, where the Bohr radius $\widetilde{a_B}=\beta a_B$ is modified compared to the noninteracting case $a_B=\sqrt{\frac{\hbar}{m\omega_0}}$. Note that for a subspace of NOs with equal occupations, one has the freedom to choose the basis states such that they match the FD states most closely. \tref{tab:01} summarizes the dominant NOs of the two-electron configurations. The states $\ket{\widetilde{0,0}}$ and $\ket{\widetilde{0,1}}$ indeed have the highest occupation numbers, and their occupations resemble the noninteraction shell fillings of FD states. In the singlet configuration, the dominant electron occupation is in $\ket{\widetilde{0,0}}$. In the triplet configuration, one electron is in $\ket{\widetilde{0,0}}$, and the other one in $\ket{\widetilde{0,1}}$. Note that a scaling factor $\beta>1$ was also used in \rcite{lowdin1956} to account for correlation effects of the helium atom.

The four-electron STQ should be described by NOs in the same way as for the two-electron STQ 
\cite{[{See also the study of the beryllium atom with NOs in }]smith1965}.
We analyze the lowest two configurations of a four-electron QD of $\hbar\omega_0=3~\text{meV}$ and $B_z=0.35~\text{T}$, which are again a singlet and a triplet. \fref{fig:02}~(b) shows the dominant NOs with its occupation numbers. In each case, we expect two electrons to be paired in $\ket{0,0}$ in a shell filling model. A wave function having this form would have $50\%$ occupation in the lowest NO. In reality, the ground state's occupation is lower, but we find that the lowest NOs for the singlet and the triplet configurations are very similar, and they are rescaled FD states $\ket{\widetilde{0,0}}$ (cf. \tref{tab:02}).

In the shell filling model, the occupations of the K shell describe the orbital properties of this STQ. For the singlet, $50\%$ of the charge should be found in $\ket{0,1}$, while for the triplet $25\%$ of the occupation should be placed each into $\ket{0,1}$ and into $\ket{0,-1}$. The NOs in \fref{fig:02}~(b) and \tref{tab:02} confirm this model with large occupations in the rescaled FD states $\ket{\widetilde{0,1}}$ and $\ket{\widetilde{0,-1}}$. Besides these dominantly occupied orbitals, we find occupations in exited NOs. Although these occupations are low, they are still higher than for the two-electron configurations.

A general implication for a computational analysis is that just a few NOs are sufficient to describe the properties of few-electron QDs. Even though each electron configuration has a distinct set of NOs, the NOs with the highest occupation numbers have a transparent interpretation. These orbitals represent the FD states with a rescaled Bohr radius. Further, the NOs with the lowest single particle energies are very similar in the singlet and triplet configurations of two-electron QDs. In the four-electron configuration, we find an identical frozen core for the singlet and triplet, and also the next higher NOs agree very well in both spin configurations. Based on the NOs that were derived in the previous calculations, we can compare the energies of all the electron configurations if we restrict the CI calculation only to the dominant NOs. \tref{tab:03} lists the energies of the two-electron and four-electron ground states for full-CI calculations and for CI calculations that are restricted to the dominant NOs. When we include the NOs with the highest occupations, very accurate descriptions of the ground state energies are obtained. For a two-electron QD, we can reproduce the results of the full-CI calculation using only four NOs with the given accuracy. The four-electron configurations are reliably described when nine NOs are included.

\section{
\label{sec:Noise}
Noise Analysis of Few-Electron Quantum Dots}

\begin{figure}
\centering
\includegraphics[width=0.49\textwidth]{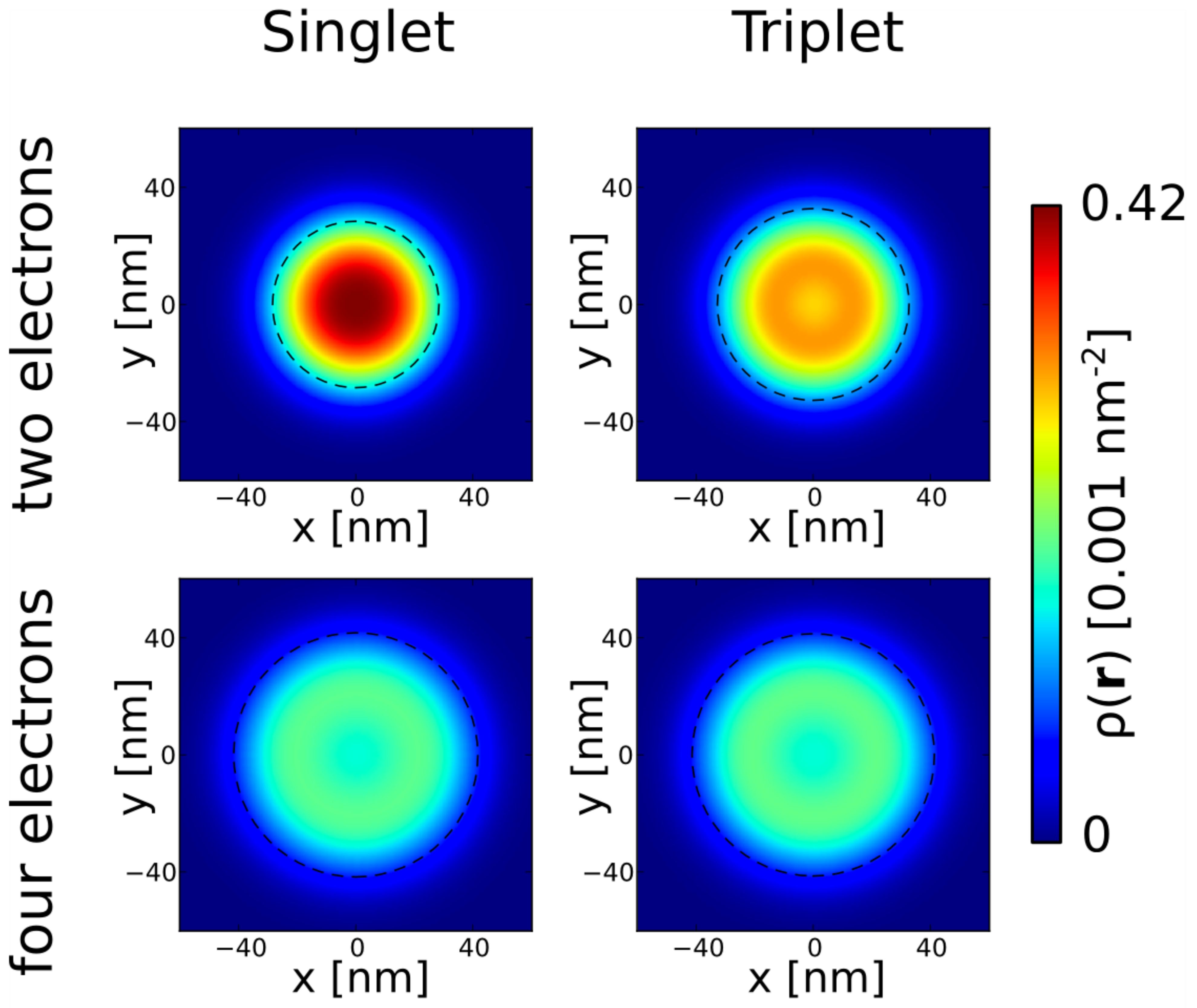}
\caption{
Probability densities $\rho\left(\bm{r}\right)$, according to \eref{eq:ProbDens}, of the lowest energy states of a two-electron QD and a four-electron QD with the confining strength $\hbar\omega_0=3~\text{meV}$ and $B_z=0.35~\text{T}$. In both cases, the ground state is in a singlet configuration and the excited state is in a triplet configuration. The dashed lines show radii where the probability densities drop to 1/e of their maximal values. The two-electron singlet has a smaller spread of the charge distribution compared to the triplet configuration. In the four-electron configuration, the probability densities are very similar for the singlet and the triplet. All properties can be explained in shell-filling description using FD states, as explained in the text.
\label{fig:03}}
\end{figure}

We start with a noise analysis for the parameters $\hbar\omega_0=3~\text{meV}$ and $B_z=0.35~\text{T}$ that were analyzed in the previous section. \fref{fig:03} shows the probability densities $\rho\left(\bm{r}\right)$ [cf. \eref{eq:ProbDens}] of the two-electron and the four-electron singlet and triplet wave functions, which are indeed the two states of lowest energies in $s_z=0$.
In the case of the two-electron QD, the singlet and the triplet probability densities are distinct from each other, which is evident in the different spread of the charge distributions. The triplet wave function has a larger spread than the singlet wave function, which is caused by the Pauli exclusion principle, which forces one electron into an orbital excited state for the triplet. In the case of the four-electron QD, the singlet and the triplet wave functions have very similar probability densities. These findings agree well with the predictions from the noninteracting orbitals of QDs which are successively filled using a shell filling method.

To characterize the degree of similarity between the singlet and the triplet configurations, we introduce the distance $D$ between the probability densities of the singlet configuration 
$\rho_S\left(\bm{r}\right)$
and the triplet configuration
$\rho_T\left(\bm{r}\right)$:
\begin{align}
D=\frac{1}{2}\int d\bm{r}
\abs{\rho_S\left(\bm{r}\right)-\rho_T\left(\bm{r}\right)}.
\label{eq:Dist}
\end{align}
$D=1$ for two nonoverlapping probability densities, and $D=0$ for two identical probability densities. For $\hbar\omega_0=3~\text{meV}$ and $B_z=0.35~\text{T}$, represented by the probability densities in \fref{fig:03}, we get $D\approx8\cdot 10^{-2}$ for the two-electron configuration and $D\approx 8\cdot 10^{-3}$ for the four-electron configuration.

We continue with a quantitative noise analysis for STQs at various magnetic fields, similar to the study in \rcite{mehl2013}. A weak single-particle perturbation $h^\prime\left(\bm{r}\right)$ renormalizes the energy difference $\delta E_{ST}$ between the single $\ket{S}$ and the triplet $\ket{T}$ states. The leading-order contribution to $\delta E_{ST}$ is given by the direct coupling of $h^\prime\left(\bm{r}\right)$ to $\ket{S}$ and $\ket{T}$:\cite{mehl2013}
\begin{align}
\delta E_{ST}=\sum_{i=1}^N\abs{
\Dirac{S}{h^\prime\left(\bm{r}_i\right)}{S}-
\Dirac{T}{h^\prime\left(\bm{r}_i\right)}{T}
}.
\end{align}
All operators $h^\prime\left(\bm{r}\right)$ that act only on the spatial part of the wave function and are velocity independent will only involve the difference in probability densities of the singlet $\rho_S\left(\bm{r}\right)$ and the triplet state $\rho_T\left(\bm{r}\right)$:
\begin{align}
\delta E_{ST}=N\int d\bm{r}~
\abs{h^\prime\left(\bm{r}\right)
\left[
\rho_S\left(\bm{r}\right)-
\rho_T\left(\bm{r}\right)
\right]
}.
\label{eq:shift_charge}
\end{align}

\begin{figure}
\centering
\includegraphics[width=0.49\textwidth]{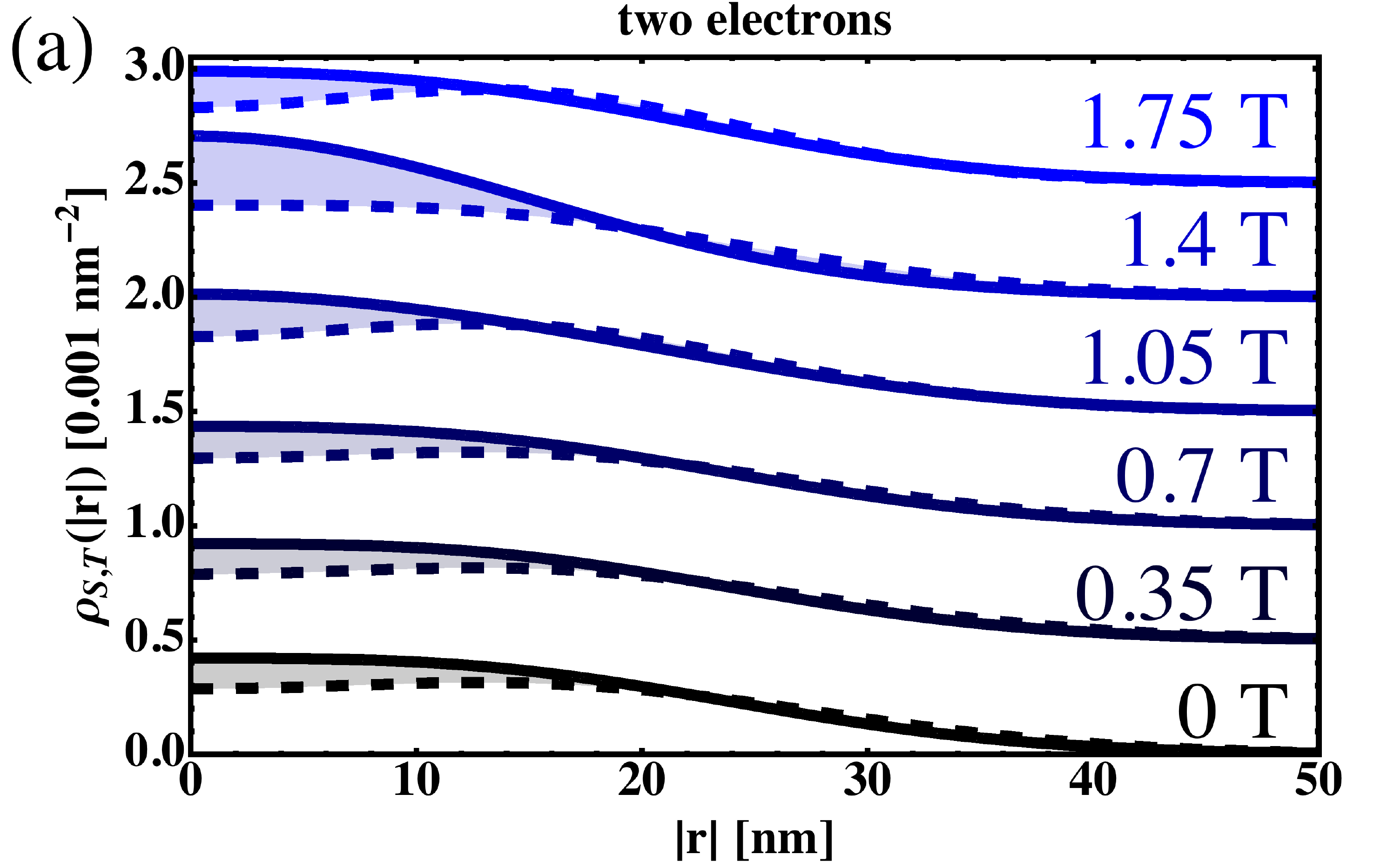}\\
\bigskip

\includegraphics[width=0.49\textwidth]{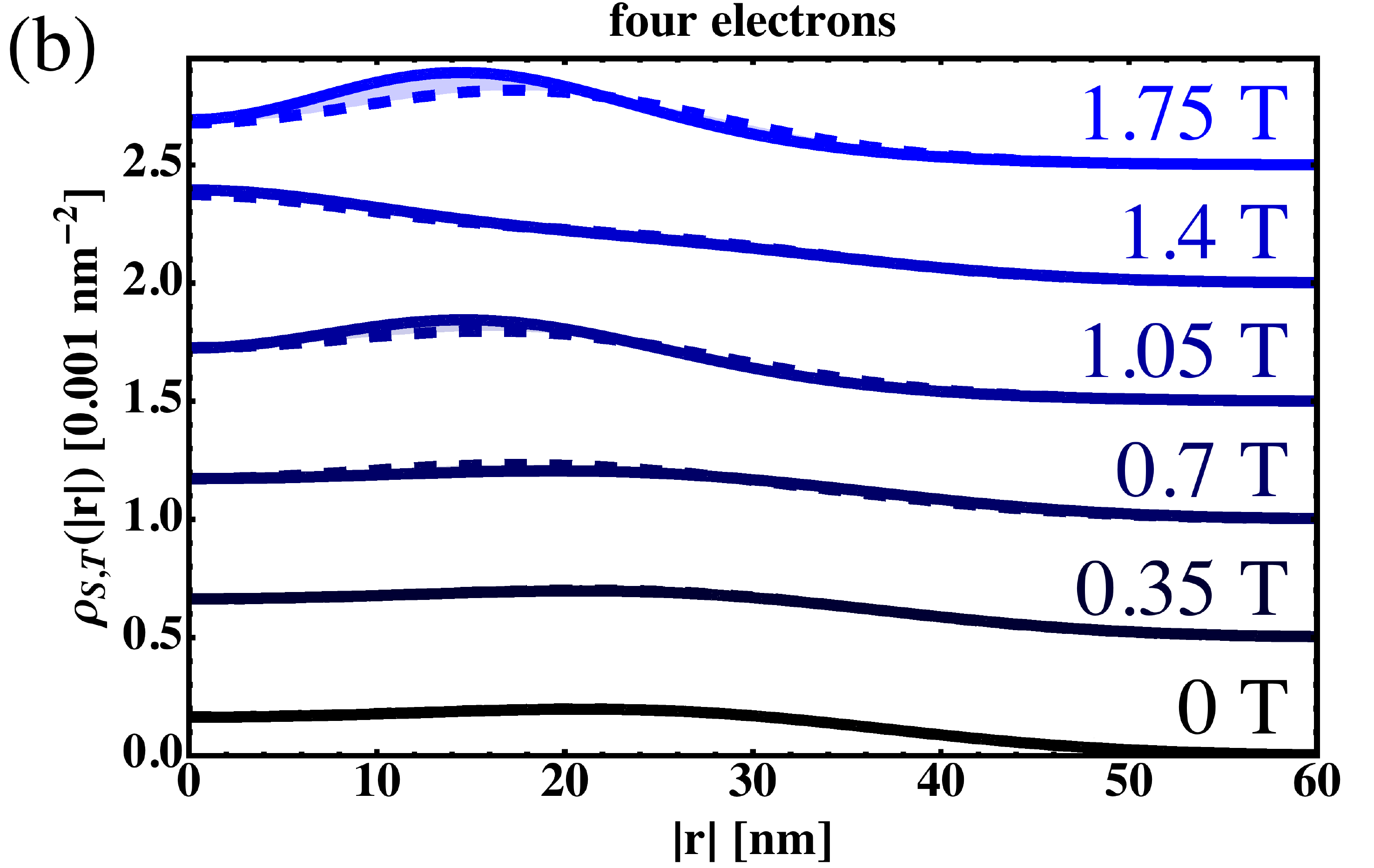}
\caption{
\label{fig:04}
Radial dependency of the singlet's [$\rho_{S}\left(\abs{\bm{r}}\right)$, solid lines] and the triplet's [$\rho_{T}\left(\abs{\bm{r}}\right)$, dashed lines] probability densities. The curves are shifted relatively to each other for different out-of-plane magnetic fields $B_z$. (a) The probability densities of the two-electron QDs are always distinct from each other. (b) For a four-electron QD, the probability densities are similar at small magnetic fields, but they differ for larger $B_z$.
}
\end{figure}

\fref{fig:04} shows the radial dependence of the probability densities of the two-electron and the four-electron configurations of a QD for different out-of-plane magnetic fields $B_z$. We find that the probability densities of the two-electron configurations are always differing more strongly from each other than the probability densities of the four-electron configurations. Especially for small $\abs{\bm{r}}$, the two-electron singlet and triplet configurations differ significantly. This finding can be understood using an orbital description for the QD electrons. In the singlet configuration, both electrons are placed into the same orbital and a high charge density is expected close to the origin. In the triplet configuration, the antisymmetry of the wave function favors the electrons to be placed further apart from each other.

In four-electron STQs, the singlet and triplet probability densities are much more similar to each other than in two-electron STQs (cf. \fref{fig:04}). Especially in small magnetic fields, $\rho_S\left(\abs{\bm{r}}\right)$ and $\rho_T\left(\abs{\bm{r}}\right)$ are matching very closely. Larger $B_z$ confine the electrons more strongly, and at some point ($>1~\text{T}$) levels of higher energy shells are lower in energy than $\ket{0,-1}$ (cf. \fref{fig:01}). As a consequence, we expect that $\rho_S\left(\abs{\bm{r}}\right)$ and $\rho_T\left(\abs{\bm{r}}\right)$ become more distinct in increasing $B_z$, which is also found when analyzing the probability densities in \fref{fig:04}.

\begin{table}
\caption{
\label{tab:04}
Probability densities of the singlet $\rho_S\left(\bm{r}\right)$ and the triplet $\rho_T\left(\bm{r}\right)$ are compared by the distance measure $D=\frac{1}{2}\int d\bm{r}
\abs{\rho_S\left(\bm{r}\right)-
\rho_T\left(\bm{r}\right)}$ and by their difference of the x component of the quadrupole moments $Q_{\psi}^{11}\equiv\int d\bm{r}~x^2\rho_{\psi}\left(\bm{r}\right)$. (a) The probability densities of two-electron STQs differ significantly. (b) For four-electron STQs, the probability densities are very similar at small $B_z$.
}
\begin{tabular}{cccc}
\hline
\hline
\multicolumn{4}{c}{(a) two-electron configuration}\\
B$_z$ [T]&D&$Q_S^{11}-Q_T^{11}$ [nm$^2$] &$\abs{\frac{Q_S^{11}-Q_T^{11}}{Q_S^{11}+Q_T^{11}}}$ [\%]\\
0 & $0.081$ & $-101$ & $8.0$\\
0.35 & $0.081$ & $-101$ & $8.0$\\
0.7 & $0.080$ & $-96$ & $7.8$\\
1.05 & $0.056$ & $-15$ & $1.3$\\
1.4 & $0.138$ & $-138$ & $13.2$\\
1.75 & $0.053$ & $6$ & $0.6$\\
\\
\multicolumn{4}{c}{(b) four-electron configuration}\\
B$_z$ [T]&D&$Q_S^{11}-Q_T^{11}$ [nm$^2$] &$\abs{\frac{Q_S^{11}-Q_T^{11}}{Q_S^{11}+Q_T^{11}}}$ [\%]\\
0 & $0.008$ & $32$ & $0.8$\\
0.35 & $0.008$ & $27$ & $0.7$\\
0.7 & $0.044$ & $136$ & $3.6$\\
1.05 & $0.053$ & $-89$ & $3.2$\\
1.4 & $0.019$ & $-4$ & $0.1$\\
1.75 & $0.087$ & $-148$ & $5.7$\\
\hline
\hline
\end{tabular}
\end{table}

Charge noise \cite{hu2006,petersson2010} dephases STQs while coupling to the charge densities of the wave functions. A QD is surrounded by many charge traps, which fluctuate between being empty or occupied. A charge trap couples to the charge distribution of the QD and dephases a STQ if the singlet's charge density differs from the triplet's charge density. We can quantify this difference using the distance measure of the probability densities $D$ according to \eref{eq:Dist}. More quantitatively, we can use the quadrupole moments of the probability density $Q^{ij}_{\psi}=\int d\bm{r}~\bm{r}_i\bm{r}_j\rho_{\psi}\left(\bm{r}\right)$ of a state $\ket{\psi}$. We approximate the shifts of the singlet-triplet energy difference $\delta E_{ST}$ that is caused by one charge trap according to \eref{eq:shift_charge}. We assume that the charge trap is some distance away from the QD and use a multipole expansion of the QD's charge distribution \cite{jackson1999}. The leading contribution to $\delta E_{ST}$ is caused by the gradient of the charge trap's electric field $\partial_{\bm{r}_i}\bm{\varepsilon}_{j}$:

\begin{align}
\delta E_{ST}
&\approx\frac{Ne}{2}\sum_{ij=1}^{3}
\abs{
\partial_{\bm{r}_i}\bm{\varepsilon}_{j}
\left(
Q^{ij}_{S}-Q^{ij}_{T}
\right)
}\\
&\approx\frac{Ne}{2}
\abs{\partial_{x} \bm{\varepsilon}_{1}+\partial_{y} \bm{\varepsilon}_{2}}
\abs{Q^{11}_{S}-Q^{11}_{T}}.
\label{eq:quadrupole}
\end{align}
Because we consider a charge distribution with rotation symmetry in the $x$-$y$ plane and identical probability
 densities of $\rho_S\left(\bm{r}\right)$ and $\rho_T\left(\bm{r}\right)$ in the $z$ direction, $Q^{11}_{S}=Q^{22}_{S}$ and $Q^{11}_{T}=Q^{22}_{T}$ are the only nontrivial components of the quadrupole tensor. The distance $d$ between the charge trap and the QD determines the prefactor in \eref{eq:quadrupole}: $\abs{\partial_{x} \bm{\varepsilon}_{1}+\partial_{y} \bm{\varepsilon}_{2}}=\frac{e^2}{4\pi\epsilon_0\epsilon_r}\frac{1}{d^3}$.

\tref{tab:04} shows that the probability densities $\rho_S\left(\bm{r}\right)$ and $\rho_T\left(\bm{r}\right)$ are always differing strongly for a two-electron QD, but they can be very similar for a four-electron QD at weak magnetic fields. For the data in \tref{tab:04}, $B_z=0.35~\text{T}$ is the optimal value to protect a STQ from charge noise because $D$ and $\abs{Q_S^{11}-Q_T^{11}}$ are both very small. We explained this finding by the filling of two NOs that have the same charge densities in \sref{sec:NO}. Note that the four-electron STQ also seems to be equally protected at zero magnetic fields and at $B=0.35~\text{T}$. STQs at $B=0$ will still have poor coherence properties because the singlet configuration is doubly degenerate. Besides the electron configuration that is equivalent to two valence electrons filled into $\ket{\widetilde{0,1}}$ [cf. \fref{fig:02}~(b)], our numerical calculation finds a state of identical energy for two valence electrons in $\ket{\widetilde{0,-1}}$.\footnote{The singlet with one electron in \protect{$\ket{\widetilde{0,1}}$} and the other one in \protect{$\ket{\widetilde{0,-1}}$} has higher energies. Even though this configuration has the same single-particle energies, the Coulomb repulsion (the Hartree contribution) is larger.}

\begin{figure}
\centering
\includegraphics[width=0.49\textwidth]{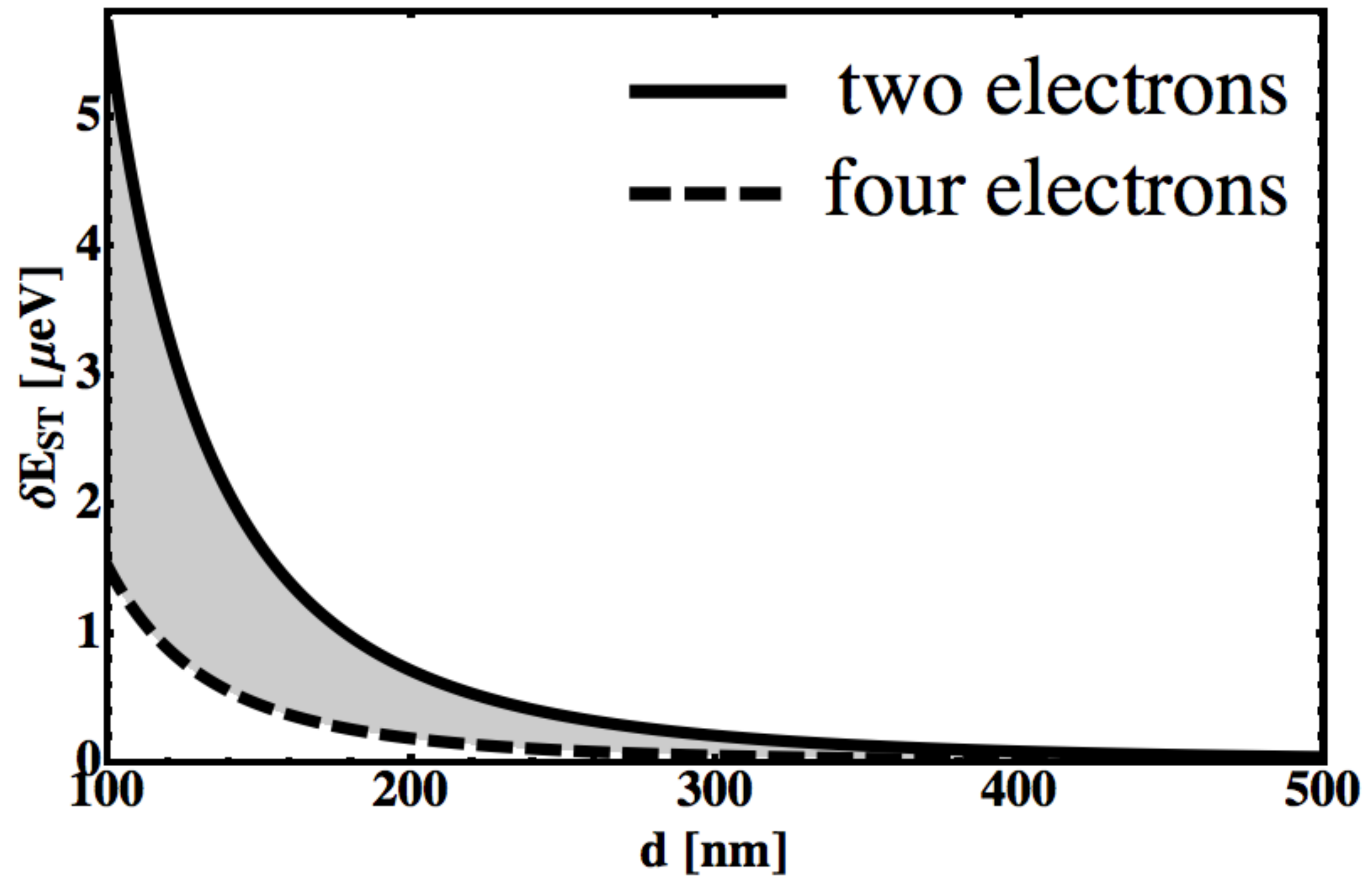}
\caption{
\label{fig:05}
Shift $\delta E_{ST}$ of the energy difference between the singlet and the triplet states of a two-electron STQ and a four-electron STQ which is caused by a charge trap that is the distance $d$ away from the QD center, as approximated by \eref{eq:quadrupole}. $\delta E_{ST}$ is nearly one order of magnitude larger for the four-electron STQ than for the two-electron STQ at $\hbar\omega_0=3~\text{meV}$ and $B_z=0.35~\text{T}$. Recall that $\delta E_{ST}=0$ for four-electron STQs in a noninteracting theory.
}
\end{figure}

The STQ encoding in four-electron QDs was considered to be optimal because the qubit states have identical electrical moments \cite{dial2013,mehl2013}. In a multipole expansion, the coupling of a charge trap to the electric quadruple moment dominates \cite{jackson1999}. \fref{fig:05} approximates the energy shifts $\delta E_{ST}$ through the coupling to the quadrupole moments from a single-site charge trap, as derived in \eref{eq:quadrupole}. We find that $\delta E_{ST}$ is nearly an order of magnitude larger for the four-electron STQ compared to the two-electron STQ. While a noninteracting theory predicts $\delta E_{ST}=0$, interacting electrons still show a strong noise protection for four-electron STQs.

\section{
\label{sec:Sum}
Discussion and Conclusion}

We have shown that the shell filling of orbitals describe the electron configurations of few-electron QDs. We explicitly studied GaAs QDs and constructed Löwdin's NOs from the interacting wave functions. Coulomb interactions rescale the radius of the FD states compared to the noninteracting orbitals of a QD. Correlations additionally require a few weakly occupied excited NOs. Nevertheless, quantitatively we find the same shell filling model for STQs with and without interactions. The immunity to charge noise of four-electron STQs, which was indicated from the noninteracting QD orbitals, stays valid for interacting QD electrons of realistic parameters. In addition, four-electron STQs at weak out-of-plane magnetic fields will be protected from all noise sources that weakly couple to the charge densities \cite{[{Note the study of the opposite limit with a strong coupling to the charge configuration in }]barnes2011}.

More generally we conclude that a CI calculation that is restricted to only a few NOs sufficiently describes the properties of few-electron QDs. Instead of deriving the NOs from a full-CI calculation, many self-consistent methods to derive NOs in CI calculations are known in quantum chemistry.\cite{mcweeny1992,szabo1996} An example is the iterative-NO method that was proposed by Bender and Davidson.\cite{bender1966} Starting from a guess for the NOs, a small set of Slater determinants is created that is used for a CI calculation. The NOs can be extracted from this calculation. Again a set of Slater determinants is created from these NOs that is used for a CI calculation. Iterating the extraction of NOs and small basis CI calculations ideally converges very rapidly. Note that the guess for the NOs does not need to resemble the noninteracting wave functions.

Especially for systems with a large influence of correlations, a full-CI calculation requires a large set of basis states to obtain an accurate description of the ground states. The iterative-NO method is one of many approaches to reduce the basis size in numerical calculations.\cite{wenzel1992,wenzel1996} While we focus only on QDs occupied with maximally four electrons and typical confining potentials, it has been shown that it is very difficult to treat weaker confinements or slightly more electrons in full-CI calculations.\cite{nielsen2013} A self-consistent NO construction would help to obtain an accurate numerical description for these QDs, and this method can also describe systems that are too big for full-CI calculations.

Our finding is also relevant to analytical descriptions of few-electron QDs. Earlier studies very often used a minimal-basis Hubbard model to describe the properties of few-electron QDs (cf., e.g., \mmrcite{mehl2013}{srinivasa2013}{kornich2014}). The chosen basis was motivated by the analysis of the noninteracting eigenfunctions of a QD potential, but it has been questionable to which degree this few-band Hubbard model represents QD wave functions when correlation effects are included. We saw that realistic QD wave functions represent the noninteracting orbital level spectrum of QDs to a high degree, and an expansion of the Hubbard model using only a few additional orbital levels can describe all correlation effects. Coulomb interactions also rescale the Bohr radius of the FD states compared to the noninteracting case. The perspective of highly accurate numerical calculations, with a transparent analytical interpretation, should further stimulate the study of NOs for QDs in the future.

\begin{acknowledgements}
We thank Maarten Wegewijs for the suggestion to use natural orbitals to describe the correction effects of quantum dot electrons. SM and DPD are grateful for support through the Alexander von Humboldt Foundation. This research has been supported by the Academy of Finland through its Centres of Excellence Program (project no. 251748).
\end{acknowledgements}

\appendix
\section{
\label{app:Numerics}
Computational Methods}

We describe the numerical analysis of $N$-electron QDs which is used in the main part of the paper. The $N$-electron configuration of QDs is described by
\begin{align}
\mathcal{H}=&\sum_{i=1}^{N}h\left(\bm{r}_i\right)+
\sum_{\av{i,j}}g\left(\bm{r}_i,\bm{r}_j\right),\\
h\left(\bm{r}\right)=&
\frac{\left(\frac{\hbar}{i}\bm{\nabla}+e\bm{A}\right)^2}{2m}+
\frac{m\omega_0}{2}\left(x^2+y^2\right),
\label{eq:HamFull2}
\\
g\left(\bm{r},\bm{r}^\prime\right)=&
\frac{e}{4\pi\epsilon_0\epsilon_r}\frac{1}{\abs{\bm{r}-\bm{r}^\prime}}.
\label{eq:HamFull3}
\end{align}
$h\left(\bm{r}\right)$ is the single-particle Hamiltonian that consists of the kinetic energy and the harmonic confining potential of a QD of the magnitude $\hbar\omega_0$. $e>0$ is the electron's charge, and $m$ is the effective mass. For the kinetic energy, we introduce the orbital effects of the magnetic field with the magnetic vector potential $\bm{A}$. Because the out-of-plane confining potential is very strong, only the out-of-plane magnetic field component is important, and we introduce its orbital effects using the symmetric gauge $\bm{A}=\frac{B_z}{2}\left(y,-x,0\right)^T$. The single-particle Hamiltonian in \eref{eq:HAMILTONIAN} can be obtained from \eref{eq:HamFull2} after some trivial rewriting. The Coulomb interaction $g\left(\bm{r},\bm{r}^\prime\right)$ in \eref{eq:HamFull3} is introduced between all the electron pairs (labeled by $\left\langle i,j\right\rangle$). $\epsilon_0$ is the dielectric constant, and $\epsilon_r$ is the relative permittivity.

For our calculations we use typical parameters of GaAs QDs. We use the effective mass $m=0.067m_e$, where $m_e$ is the electron mass, and $\epsilon_r=12.7$ is the relative permittivity. Typical GaAs QDs have $\hbar\omega_0=3~\text{meV}$. The Hamiltonian is discretized using a grid of Gaussian basis functions \cite{nielsen2010}, and the full-CI calculation is done using a basis of $50$ single-particle eigenstates of the Hamiltonian from \eref{eq:HamFull2} with the lowest energies. This number of basis states was sufficient for the convergence of the full-CI calculations in earlier calculations of two-electron STQs \cite{hiltunen2014,hiltunen2015}, and we can achieve similar convergence with the four-electron STQs (an energy cutoff similar to \rcite{barnes2011} was used here to restrict the size of the CI basis).

\section{
\label{app:RedDens}
Method of Reduced Density Matrices}

This section describes how reduced density matrices describe correlated quantum systems \cite{lowdin1955-1,*lowdin1955-2,*lowdin1955-3}. We define an $N$-electron Hamiltonian
\begin{widetext}
\begin{align}
\label{eq:Ham}
\mathcal{H}=&\sum_{i=1}^{N}h\left(\bm{r}_i\right)+
\underbrace{\sum_{\av{i,j}}g\left(\bm{r}_i,\bm{r}_j\right)}_{
\frac{1}{2}\sum_{ij}^\prime g\left(\bm{r}_i,\bm{r}_j\right)
}
\\\nonumber
=&\int d\bm{r}~
\hat{\Psi}^\dagger\left(\bm{r}\right)
h\left(\bm{r}\right)
\hat{\Psi}\left(\bm{r}\right)
+
\frac{1}{2}
\int d\bm{r}\int d\bm{r}^\prime~
\hat{\Psi}^\dagger\left(\bm{r}\right)
\hat{\Psi}^\dagger\left(\bm{r}^\prime\right)
g\left(\bm{r},\bm{r}^\prime\right)
\hat{\Psi}\left(\bm{r}^\prime\right)
\hat{\Psi}\left(\bm{r}\right)
\\\nonumber
=&\sum_{n\sigma,m\sigma^\prime}
h^{n\sigma}_{m\sigma^\prime}
\cre{n\sigma}\ann{m\sigma^\prime}
+
\frac{1}{2}\sum_{
\begin{array}{l}
\scriptstyle n\sigma,m\sigma^\prime\\
\scriptstyle k\sigma^{\prime\prime},l\sigma^{\prime\prime\prime}
\end{array}}
g^{n\sigma,m\sigma^\prime}_{k\sigma^{\prime\prime},l\sigma^{\prime\prime\prime}}
\cre{n\sigma}\cre{m\sigma^\prime}\ann{l\sigma^{\prime\prime\prime}}\ann{k\sigma^{\prime\prime}},
\end{align}
\end{widetext}
with the single-particle Hamiltonian $h\left(\bm{r}\right)$ and the interaction Hamiltonian $g\left(\bm{r},\bm{r}^\prime\right)$. $\av{i,j}$ singles out all the pairs of the $N$ particles, and $\sum_{ij}^\prime$ sums over all the particles except identical ones. We introduce also the notations in second quantization. $\hat{\Psi}^{\left(\dagger\right)}\left(\bm{r}\right)$ is the field annihilation (creation) operator of a particle at position $\bm{r}$. We introduced a basis $\left\{\ket{n\sigma}\right\}$ of orthogonal and normalized wave functions. $\hat{c}^{\left(\dagger\right)}_{n\sigma}$ is the annihilation (creation) operator of a particle in the state $\ket{n\sigma}$. $h^{n\sigma}_{m\sigma^\prime}
=\Dirac{n\sigma}{h}{m\sigma^\prime}$ and $g^{n\sigma,m\sigma^\prime}_{k\sigma^{\prime\prime},l\sigma^{\prime\prime\prime}}
=\Dirac{n\sigma,m\sigma^\prime}{g}{k\sigma^{\prime\prime},l\sigma^{\prime\prime\prime}}$ are the spectral representations of the Hamiltonians in the basis $\left\{\ket{n\sigma}\right\}$.

Löwdin introduced reduced density matrices to describe $N$-electron wave functions\cite{lowdin1955-1,*lowdin1955-2,*lowdin1955-3}. We define the first-order density matrix:
\begin{align}
\varrho\left(\bm{r},\bm{r}^\prime\right)
=&N\int
d\bm{r}_2\cdot\dotsc\cdot d\bm{r}_N
\\\nonumber
&\times
\psi^*\left(\bm{r},\bm{r}_2,\dotsc, \bm{r}_N\right)
\psi\left(\bm{r}^\prime,\bm{r}_2,\dotsc, \bm{r}_N\right).
\label{eq:First-Order}
\end{align}
$\varrho\left(\bm{r},\bm{r}\right)d\bm{r}$ describes the probability that one particle is found in the area $d\bm{r}$ around the point $\bm{r}$. The first-order density matrix can be conveniently rewritten in second quantization:
\begin{align}
\varrho\left(\bm{r},\bm{r}^\prime\right)&=
\Dirac{\psi}{
\hat{\Psi}^\dagger\left(\bm{r}\right)
\hat{\Psi}\left(\bm{r}^\prime\right)
}{\psi}\\
&\nonumber=
\sum_{n\sigma,m\sigma^\prime}
\varrho^{n\sigma}_{m\sigma^\prime}
\phi_{n\sigma}^*\left(\bm{r}\right)
\phi_{m\sigma^\prime}\left(\bm{r}^\prime\right).
\end{align}
$\varrho^{n\sigma}_{m\sigma^\prime}
=\Dirac{\psi}{
\hat{c}^{\dagger}_{n\sigma}
\hat{c}_{m\sigma^\prime}
}{\psi}$ is the spectral decomposition of the first-order density matrix in terms of the wave functions $\phi_{n\sigma}\left(\bm{r}\right)=\product{\bm{r}}{n\sigma}$.

Similarly, one defines the second-order density matrix:
\begin{align}
\pi&\left(\overline{\bm{r}},\overline{\bm{r}}^\prime,\bm{r},\bm{r}^\prime\right)
=N\left(N-1\right)
\int d\bm{r}_3\cdot\dotsc\cdot d\bm{r}_N
\\\nonumber
&\times
\psi^*\left(\overline{\bm{r}},\overline{\bm{r}}^\prime,\bm{r}_3,\dotsc, \bm{r}_N\right)
\psi\left(\bm{r},\bm{r}^\prime,\bm{r}_3,\dotsc, \bm{r}_N\right).
\end{align}
$\pi\left(\bm{r},\bm{r}^\prime,\bm{r},\bm{r}^\prime\right)d\bm{r}d\bm{r}^\prime$ describes the probability that one particle is found in the volume $d\bm{r}$ around $\bm{r}$ and one particle is found in the volume $d\bm{r}^\prime$ around $\bm{r}^\prime$. Also the second-order density matrix can be rewritten in second quantization:
\begin{widetext}
\begin{align}
\pi\left(\overline{\bm{r}},\overline{\bm{r}}^\prime,\bm{r},\bm{r}^\prime\right)&=
\Dirac{\psi}{
\hat{\Psi}^\dagger\left(\overline{\bm{r}}\right)
\hat{\Psi}^\dagger\left(\overline{\bm{r}}^\prime\right)
\hat{\Psi}\left(\bm{r}^\prime\right)
\hat{\Psi}\left(\bm{r}\right)
}{\psi}\\
&\nonumber=
\sum_{
\begin{array}{l}
\scriptstyle n\sigma,m\sigma^\prime\\
\scriptstyle k\sigma^{\prime\prime},l\sigma^{\prime\prime\prime}
\end{array}}
\pi^{n\sigma,m\sigma^\prime}_{k\sigma^{\prime\prime},l\sigma^{\prime\prime\prime}}
\phi_{n\sigma}^*\left(\overline{\bm{r}}\right)
\phi_{m\sigma^\prime}^*\left(\overline{\bm{r}}^\prime\right)
\phi_{k\sigma^{\prime\prime}}\left(\bm{r}\right)
\phi_{l\sigma^{\prime\prime\prime}}\left(\bm{r}^\prime\right),
\end{align}
\end{widetext}
with the spectral decomposition $
\pi^{n\sigma,m\sigma^\prime}_{k\sigma^{\prime\prime},l\sigma^{\prime\prime\prime}}=
\Dirac{\psi}{
\hat{c}^{\dagger}_{n\sigma}
\hat{c}^{\dagger}_{m\sigma^\prime}
\hat{c}_{l\sigma^{\prime\prime\prime}}
\hat{c}_{k\sigma^{\prime\prime}}
}{\psi}$ of the density matrix.

It is very convenient to calculate expectation values using the reduced density matrix notation. For example, the expectation values of the single-particle Hamiltonian of \eref{eq:Ham},
\begin{align}
\av{\sum_ih\left(i\right)}
=\sum_{n\sigma,m\sigma^\prime}
\varrho^{n\sigma}_{m\sigma^\prime}
h^{n\sigma}_{m\sigma^\prime},
\end{align}
and the two-particle Hamiltonian,
\begin{align}
\av{\sum_{i,j}{}^\prime g\left(i,j\right)}=
\sum_{
\begin{array}{l}
\scriptstyle n\sigma,m\sigma^\prime\\
\scriptstyle k\sigma^{\prime\prime},l\sigma^{\prime\prime\prime}
\end{array}}
\pi^{n\sigma,m\sigma^\prime}_{k\sigma^{\prime\prime},l\sigma^{\prime\prime\prime}}
g^{n\sigma,m\sigma^\prime}_{k\sigma^{\prime\prime},l\sigma^{\prime\prime\prime}},
\end{align}
are now easy to calculate.

\bibliography{library}
\end{document}